\documentclass[aps,pra,twocolumn,showpacs,superscriptaddress]{revtex4-1}
\usepackage{bm}
\usepackage{mathrsfs}
\usepackage{amsmath}
\usepackage{amssymb}
\usepackage{graphicx}
\usepackage{amsfonts}
\usepackage{amsthm}
\usepackage{color}
\usepackage{dcolumn}
\usepackage{txfonts}

\begin{document}

\title{Insights into Phase Transitions and Entanglement from Density Functional Theory}
\author{Bo-Bo Wei}
\affiliation{School of Physics and Energy, Shenzhen University, Shenzhen, China}

\begin{abstract}
Density functional theory has made great success in solid state physics, quantum chemistry and in computational material sciences. In this work we show that density functional theory could shed light on phase transitions and entanglement at finite temperatures. Specifically, we show that the equilibrium state of an interacting quantum many-body system which is in thermal equilibrium with a heat bath at a fixed temperature is a universal functional of the first derivatives of the free energy with respect to temperature and other control parameters respectively. This insight from density functional theory enables us to express the average value of any physical observable and any entanglement measure as a universal functional of the first derivatives of the free energy with respect to temperature and other control parameters. Since phase transitions are marked by the nonanalytic behavior of free energy with respect to control parameters, the physical quantities and entanglement measures may present nonanalytic behavior at critical point inherited from their dependence on the first derivative of free energy. We use an experimentally realizable model to demonstrate the idea. These results give new insights for phase transitions and provide new profound connections between entanglement and phase transition in interacting quantum many-body physics.
\end{abstract}


\maketitle

\section{Introduction}

The electronic density functional theory (DFT) developed by Hohenberg and Kohn\cite{HK1964} and Kohn and Sham \cite{KS1965} in 1964-1965 has shown tremendous success in solid state physics, quantum chemistry and in computational material sciences \cite{DFT2015a,DFT2015b}. The central idea of DFT is a transformation of the dependence of the properties of a system of interacting many-body system on their single-particle potential to a dependence on the ground state density, which provides a practical method to calculate ground state properties of electronic systems\cite{DFT2015a,DFT2015b}. David Mermin generalized the DFT formalism to calculate finite temperature properties \cite{Merin1965} and Runge and Gross extended DFT further to calculate time-dependent properties and hence the excited state properties of electronic systems \cite{Runge1984}.

Phase transitions are one of the most intriguing phenomena in many-body physics because it indicates emergence of new states of matter. In recent years, there are a great deal of interest in studying the relations between entanglement and phase transitions in many-body systems \cite{Vedral2008}. An inspiring result on the relations between entanglement and quantum phase transitions is from DFT perspective \cite{Wu2005}. It was also shown that DFT provides new insights for quantum phase transitions\cite{Nagy2013a,Nagy2013b}. However, any realistic experiments has always been performed at finite temperatures, it is therefore desirable to connect the phase transitions and entanglement at finite temperatures.

In this work we show that density functional theory provides insights for finite temperature phase transitions. We prove that the equilibrium state of a quantum many-body system, which is in thermal equilibrium with a heat bath, is a universal functional of the first derivative of the free energy with respect to the control parameters. This finding explains how the non-analytic behavior of free energy at critical point affects the expectation values of physical observable at phase transition point. Since entanglement in quantum many-body systems is a functional of expectation value of observable, the finding also introduces a direct link between entanglement and the first derivatives of free energy, leading to a deep connection between entanglement and phase transitions. The organization of this paper is as follows. We prove a few theorems for establishing the relation between thermodynamic equilibrium state and first derivatives of free energy from density functional theory in Sec.~II. In Sec.~III, we prove a theorem for connecting thermal phase transitions and entanglement. In Sec.~VI we study an experimentally relevant model to demonstrate our central ideas and finally we give a summary.

\section{Insight into Phase Transitions from Density Functional Theory}
Let us consider an interacting quantum many-body system with Hamiltonian
\begin{equation}\label{ham}
H(\{\lambda_i\})=H_0+\sum_{i=1}^K\lambda_iH_i,
\end{equation}%
where $\{\lambda_i\}=(\lambda_1,\lambda_2,\cdots,\lambda_K)$ are the control parameters of the system. We shall work in the canonical ensemble, where the many-body system is in contact with a heat bath at a fixed temperature and finally they reach thermal equilibrium. The thermodynamic equilibrium state of the many-body system can be obtained from maximizing the entropy \cite{Jaynes1957a,Jaynes1957b}
\begin{equation}
S=-\text{Tr}\Big[\rho\ln\rho\Big],
\end{equation}
under two constraints. The first constraint is that the average energy of the system is fixed,
\begin{eqnarray}
\text{Tr}[\rho H]&=&\langle E\rangle,
\end{eqnarray}
which is because the system is in thermal equilibrium with the heat bath and there is no macroscopic energy flow between them. The second constraints is that the density matrix is normalized, namely
\begin{eqnarray}
\text{Tr}[\rho]&=&1.
\end{eqnarray}
By maximizing the entropy under above two constraints, one obtain the equilibrium state of the system with density matrix,
\begin{eqnarray}\label{dm}
\rho_0\equiv \frac{e^{-\beta H(\lambda_1,\cdots,\lambda_K)}}{Z(T,\lambda_1,\cdots,\lambda_K)},
\end{eqnarray}
where $\beta\equiv1/T$ and $Z(T,\lambda_1,\cdots,\lambda_K)\equiv\text{Tr}[e^{-\beta H(\lambda_1,\cdots,\lambda_K)}]$ being the canonical partition function. We take the Boltzmann constant $k_B\equiv1$. One can see from Equation \eqref{dm} that the thermal equilibrium state is fully characterized by the set of parameters $(T,\lambda_1,\lambda_2,\cdots,\lambda_K)$. We shall prove that the reverse is also true by establishing the first theorem:\\
\textbf{Theorem 1:} There is a one-to-one map between the thermodynamic equilibrium state of a quantum many-body system with Hamiltonian $H=H_0+\sum_{i=1}^K\lambda_iH_i$, which is in thermal equilibrium state with a heat bath at temperature $T$, and the set of control parameters $(T,\lambda_1,\lambda_2,\cdots,\lambda_K)$.\\
\emph{Proof}: In the above we show that the equilibrium state is described by the density matrix $\rho_0=e^{-\beta H}/Z(T,\lambda_1,\lambda_2,\cdots,\lambda_K)$, which means that the equilibrium state is determined by the control parameters $(T,\lambda_1,\lambda_2,\cdots,\lambda_K)$. Now we have to prove that the equilibrium state $\rho_0$ determines these control parameters. That means only one set of control parameters corresponds to the equilibrium state $\rho_0$. The proof is done by reductio and absurdum. We assume that there are two different sets of control parameters $(T,\lambda_1,\lambda_2,\cdots,\lambda_K)$ and $(T',\lambda_1',\lambda_2',\cdots,\lambda_K')$ corresponds to the same equilibrium state $\rho_0$. Then we have
\begin{eqnarray}
\text{Tr}[\rho_0 H(\lambda_1,\lambda_2,\cdots,\lambda_K)]&=&\langle E\rangle,\label{theorem1a}\\
\text{Tr}[\rho_0H(\lambda_1',\lambda_2',\cdots,\lambda_K')]&=&\langle E'\rangle\label{theorem1b}.
\end{eqnarray}
Note that the information of temperature is embodied by the average energy of the system. Subtracting Equation \eqref{theorem1a} from \eqref{theorem1b}, we get
\begin{eqnarray}
\sum_{i=1}^K(\lambda_i-\lambda_i')\text{Tr}[\rho_0H_i]+\langle E\rangle-\langle E'\rangle=0.
\end{eqnarray}
Thus we have
\begin{eqnarray}
\lambda_i&=&\lambda_i', i=1,2,\cdots,K.\\
\langle E\rangle&=&\langle E'\rangle. \label{theorem1c}
\end{eqnarray}
Equation \eqref{theorem1c} means that $T=T'$. This contradicts the assumption. Thus we proved Theorem I. Theorem 1 means that the thermal equilibrium state is uniquely fixed by the control parameters, namely
\begin{eqnarray}
\rho_0\Longleftrightarrow \Big(T,\lambda_1,\lambda_2,\cdots,\lambda_K\Big).
\end{eqnarray}
From the equilibrium density matrix $\rho_0$, we can evaluate all physical quantities, in particular, the quantities conjugate to the control parameters. For example the entropy, which is conjugate to temperature, is given by 
\begin{eqnarray}
S_0&=&-\text{Tr}[\rho_0\ln\rho_0].
\end{eqnarray}
The density $\langle H_i\rangle$ which is conjugate to $\lambda_i$ is 
\begin{eqnarray}
\langle H_i\rangle&=&\text{Tr}[\rho_0H_i], i=1,2,\cdots,K.
\end{eqnarray}
Theorem 1 implies that the control parameters $(T,\lambda_1,\lambda_2,\cdots,\lambda_K)$ determine the densities $(S_0,\langle H_1\rangle,\langle H_2\rangle,\cdots,\langle H_K\rangle)$. We shall prove the reverse is also true in Theorem 2.

To establish Theorem 2, we record a minimum property of the free energy at finite temperature in analogous to that of the ground state energy. If we define a free energy functional
\begin{eqnarray}\label{a0}
F[T,\{\lambda_i\},\rho]=\text{Tr}\Big[\rho\Big(H(\{\lambda_i\})+T\ln\rho\Big)\Big],
\end{eqnarray}
then this functional satisfies
\begin{eqnarray}\label{a1}
F[T,\{\lambda_i\},\rho]>F[T,\{\lambda_i\},\rho_0], \rho\neq\rho_0
\end{eqnarray}
for all positive definite density matrices $\rho$ with unit trace and $\rho_0$ is the equilibrium state corresponds to the control parameters $(T,\{\lambda_i\})$. In the Appendix A we prove that $F[T,\{\lambda_i\},\rho]$ is bounded below by $F[T,\{\lambda_i\},\rho_0]$ for general quantum many-body systems. Given this property, we can prove \\
\textbf{Theorem 2:} For an interacting quantum many-body system with Hamiltonian, $H=H_0+\sum_{i=1}^K\lambda_i H_i$, which is in thermal equilibrium with a heat bath at fixed temperature $T$, there is a one-to-one map between the set of control parameters $\{T,\lambda_1,\cdots,\lambda_K\}$ and thermal equilibrium densities, $\{S_0,\langle H_1\rangle,\cdots,\langle H_K\rangle\}$.\\
\emph{Proof}: For simplicity of notation, we denote the set of control parameters by $\Lambda\equiv\{T,\lambda_1,\cdots,\lambda_K\}$.  Let us consider two different sets of parameters $\Lambda\neq\Lambda'$ and assume that their corresponding equilibrium density matrices are $\rho_0$ and $\rho_0'$, respectively. From their density matrices, we can obtain the sets for thermal densities $D_{\Lambda}\equiv\{S_0,\langle H_1\rangle,\cdots,\langle H_K\rangle\}$ and $D_{\Lambda'}\equiv\{S_0',\langle H_1\rangle',\cdots,\langle H_K\rangle'\}$ where $S_0=-\text{Tr}[\rho_0\ln\rho_0]$, $\langle H_i\rangle=\text{Tr}[\rho_0H_i],i=1,2,\cdots,K$ and $S_0'=-\text{Tr}[\rho_0'\ln\rho_0']$, $\langle H_i\rangle'=\text{Tr}[\rho_0'H_i],i=1,2,\cdots,K$. We assume two different sets of control parameters $\Lambda$ and $\Lambda'$ produce the same set for thermal equilibrium densities, namely $D_{\Lambda}=D_{\Lambda'}$. According to the minimum property of free energy, Equation\eqref{a1}, we have
\begin{eqnarray}
F[\Lambda,\rho_0]&=&\text{Tr}\Big[\rho_0\Big(H_0+\sum_{i=1}^K\lambda_i H_i+T\ln\rho_0\Big)\Big],\label{theorem2a}\\
&<&F[\Lambda,\rho_0'],\label{theorem2aa} \\
&=&\text{Tr}\Big[\rho_0'\Big(H_0+\sum_{i=1}^K\lambda_i H_i+T\ln\rho_0'\Big)\Big],\label{theorem2b}\\
&=&F[\Lambda',\rho_0']+\sum_{i=1}^K(\lambda_i-\lambda_i')\langle H_i\rangle'+(T-T')S_0'\label{theorem2c},
\end{eqnarray}
so that
\begin{eqnarray}\label{b2}
F[\Lambda,\rho_0]<F[\Lambda',\rho_0']+\sum_{i=1}^K(\lambda_i-\lambda_i')\langle H_i\rangle'+(T-T')S_0'.
\end{eqnarray}
But the reasoning of Equation\eqref{theorem2a} to Equation\eqref{theorem2c} remains valid when the set for parameters $\Lambda$ and $\Lambda'$ are interchanged, yielding
\begin{eqnarray}\label{b3}
F[\Lambda',\rho_0']<F[\Lambda,\rho_0]+\sum_{i=1}^K(\lambda_i'-\lambda_i)\langle H_i\rangle+(T'-T)S_0.
\end{eqnarray}
Adding Equation\eqref{b2} and Equation\eqref{b3} and using the assumptions lead to the contradiction
\begin{eqnarray}
F[\Lambda,\rho_0]+F[\Lambda',\rho_0']<F[\Lambda,\rho_0]+F[\Lambda',\rho_0'],
\end{eqnarray}
and therefore a set of control parameters $\Lambda=\{T,\lambda_1,\cdots,\lambda_K\}$ can only result in a set of thermal densities $D_{\Lambda}=\{S_0,\langle H_1\rangle,\cdots,\langle H_K\rangle\}$, i.e. the control parameters are unique function of the equilibrium densities.  Theorem 2 is proved. Theorem II means that
\begin{eqnarray}
\Big(T,\lambda_1,\lambda_2,\cdots,\lambda_K\Big)\Longleftrightarrow \Big(S_0,\langle H_1\rangle,\langle H_2\rangle,\cdots,\langle H_K\rangle\Big).
\end{eqnarray}
An immediate consequence of Theorem 2 is that we can use the thermal densities instead of control parameters as a fundamental variable of the equilibrium state, $\rho_0(S_0,\langle H_1\rangle,\cdots,\langle H_K\rangle)$. In real applications, it is always useful to vary only one of the control parameters while keeping all the others fixed. Then Theorem 2 actually implies that the following one-to-one correspondence relations,
\begin{eqnarray}
T &\Longleftrightarrow& S_0,\\
\lambda_i &\Longleftrightarrow& \langle H_i\rangle,\ \text{for} \ i=1,2,\cdots,K.
\end{eqnarray}
Now we are ready to establish the following theorem:\\
\textbf{Theorem 3:} If an interacting many-body system with Hamiltonian $H(\lambda_1,\cdots,\lambda_K)=H_0+\sum_{i=1}^K\lambda_i H_i$ is in thermal equilibrium with a heat bath at temperature $T$, then the equilibrium density matrix of the many-body system is a unique functional of the first order derivatives of the free energy with respect to control parameters, namely
\begin{eqnarray}\label{b4}
\rho_0=\rho_0\Big(\frac{\partial F}{\partial T},\frac{\partial F}{\partial\lambda_1},\cdots,\frac{\partial F}{\partial\lambda_K}\Big).
\end{eqnarray}
\emph{Proof}: According to Theorem 2, we have one-to-one correspondence relations between thermal densities and the corresponding control parameters. Thus we can express the control parameter as function of their conjugate density, namely
\begin{eqnarray}
T&=&f_0(S),\\
\lambda_i&=&f_i\Big(\langle H_i\rangle\Big), \ i=1,2,\cdots,K,
\end{eqnarray}
where $f_0,f_1,f_2,\cdots,f_K$ are some unknown functions. Because the thermal densities are all first derivatives of free energy, for example, entropy is the first derivative of free energy with respect to temperature $S=-\frac{\partial F}{\partial T}$ and $\langle H_i\rangle=\frac{\partial F}{\partial\lambda_i},i=1,2,\cdots,K$, this means that
\begin{eqnarray}
T&=&f_0\Big(\frac{\partial F}{\partial T}\Big),\label{theorem3a}\\
\lambda_i&=&f_i\Big(\frac{\partial F}{\partial\lambda_i}\Big), \ i=1,2,\cdots,K.\label{theorem3b}
\end{eqnarray}
In addition, Theorem 1 tells us that the equilibrium state is uniquely fixed by the control parameters, i.e. $\rho_0(T,\lambda_1,\lambda_2,\cdots,\lambda_K)$. Combing with Equations \eqref{theorem3a} and \eqref{theorem3b} with Theorem 1, we proved Theorem 3. Theorem 3 means that we can use the first derivative of free energy with respect to control parameters as a fundamental variable.

According to Theorem 3, the thermal expectation value of any observable $A$ of the many-body system is given by
\begin{eqnarray}\label{average}
\langle A\rangle=\text{Tr}\Big[A\rho_0\Big(\frac{\partial F}{\partial T},\frac{\partial F}{\partial\lambda_1},\cdots,\frac{\partial F}{\partial\lambda_K}\Big)\Big].
\end{eqnarray}
Phase transitions are marked by the discontinuity of the free energy with respect to temperature $T$ or other control parameters $\lambda_i,i=1,2,\cdots,K$: \\
(1). For first order phase transitions, the first order derivatives of the free energy, such as $\partial_T F$ or $\partial_{\lambda_i} F,i=1,2\cdots$, is discontinuous across a phase boundary, then expectation value of an arbitrary observable $A$ also presents discontinuity across phase boundary from Equation\eqref{average}. \\
(2).~For second order phase transitions, the first derivatives of free energy with respect to parameters are continuous and the second order derivatives of the free energy, such as $\partial_{T}^2F$ and $\partial_{\lambda_i}^2F$, are discontinuous or diverge and we should examine the derivative of the observable $A$, which is
\begin{eqnarray}\label{b5}
\frac{\partial\langle A\rangle}{\partial T}=\text{Tr}\Big[A\frac{\partial\rho_0}{\partial S_0}\Big]\times\frac{\partial S_0}{\partial T}=-\text{Tr}\Big[A\frac{\partial\rho_0}{\partial S_0}\Big]\times\frac{\partial^2F}{\partial T^2}.
\end{eqnarray}
\begin{eqnarray}\label{b6}
\frac{\partial\langle A\rangle}{\partial\lambda_i}=\text{Tr}\Big[A\frac{\partial\rho_0}{\partial \langle H_i\rangle}\Big]\times\frac{\partial\langle H_i\rangle}{\partial \lambda}&=&\text{Tr}\Big[A\frac{\partial\rho_0}{\partial\langle H_i\rangle}\Big]\times\frac{\partial^2F}{\partial \lambda_i^2},\\
i&=&1,2,\cdots,K.\nonumber
\end{eqnarray}
Equation \eqref{b5} and\eqref{b6} show that the first order derivative of any observable $A$ with respect to control parameters is proportional to the second order derivative of free energy with respect to the same control parameter. While for second order thermal phase transitions, $\partial_{T}^2F$ and $\partial_{\lambda}^2F$ are discontinuous or diverge, thus the first derivative $\partial_T \langle A\rangle$ and $\partial_{\lambda}\langle A\rangle$ present discontinuous or divergence at second order phase transition point. Analogously, one needs to detect $(n-1)$-th order derivative of physical quantity for $n-$th order phase transitions.

\section{Relations between entanglement and phase transitions from density functional theory}
Entanglement is a unique feature in quantum mechanics. It was found that entanglement measures present non-analytic behavior at phase transition point \cite{Vedral2008}. Wu \emph{et. al.} \cite{Wu2005} found that density functional theory provide intriguing relationship between entanglement and quantum phase transitions. Now we show that density functional theory provides relations between entanglement and phase transitions in many-body systems at any temperatures. We can prove the following theorem:\\
\textbf{Theorem 4. } Any finite temperature entanglement measure $M$ can be expressed as a unique functional of the set of first derivatives of the free energy:
\begin{eqnarray}
M&=&M\Big(S_0,\langle H_1\rangle,\langle H_2\rangle,\cdots,\langle H_K\rangle\Big)\\
&=&M\Big(\frac{\partial F}{\partial T},\frac{\partial F}{\partial\lambda_1},\frac{\partial F}{\partial\lambda_2},\cdots,\frac{\partial F}{\partial\lambda_K}\Big).
\end{eqnarray}
\emph{Proof}: The proof follows from the fact that, according to Theorem 1 and 2, the thermal equilibrium density matrix is a unique functional of $S_0,\langle H_1\rangle,\langle H_2\rangle,\cdots,\langle H_K\rangle$ and also $\rho_0$ provides the complete information of the thermal equilibrium state, everything else is a unique functional of $(S_0,\langle H_1\rangle,\langle H_2\rangle,\cdots,\langle H_K\rangle)$. Formally let us consider an $n$-partite entanglement in spin-1/2 systems. For other cases, the proof can be generalized immediately. First of all any entanglement measure of $n$ qubits is always a function of the matrix elements of the reduced density matrix of these qubits, $\rho_{12\cdots n}$: $M(\rho_{12\cdots n})$. For spin-1/2 systems, the $n$-body reduced density matrix can be written as
\begin{eqnarray}
\rho_{12\cdots n}=\sum_{a_1,a_2,\cdots,a_n=0,x,y,z}C_{a_1a_2,\cdots,a_n}\sigma_1^{a_1}\sigma_2^{a_2}\cdots\sigma_{n}^{a_n},
\end{eqnarray}
with
\begin{eqnarray}
C_{a_1a_2,\cdots,a_n}&=&\text{Tr}_{12\cdots n}[\rho_{12\cdots n}\sigma_1^{a_1}\sigma_2^{a_2}\cdots\sigma_n^{a_n}],\\
&=&\text{Tr}[\rho_0\sigma_1^{a_1}\sigma_2^{a_2}\cdots\sigma_n^{a_n}],\\
&=&\langle\sigma_1^{a_1}\sigma_2^{a_2}\cdots\sigma_n^{a_n}\rangle,
\end{eqnarray}
where $a_1,a_2,\cdots,a_n=0,x,y,z$ with $\sigma^0=I$. Thus $M=M\Big(\langle\sigma_1^{a_1}\sigma_2^{a_2}\cdots\sigma_n^{a_n}\rangle\Big)$. According to Theorem 2, thermal expectation value of any observable can be taken as a functional of $(S_0,\langle H_1\rangle,\langle H_2\rangle,\cdots,\langle H_K\rangle)$. Therefore $M=M\Big(S_0,\langle H_1\rangle,\cdots,\langle H_K\rangle\Big)=M\Big(\frac{\partial F}{\partial T},\frac{\partial F}{\partial\lambda_1},\cdots,\frac{\partial F}{\partial\lambda_K}\Big)$. Theorem 4 is proved.

One can immediately see that for first order phase transition, where the first derivative of free energy is discontinuous, the entanglement measure presents non-analytic behavior. For second order phase transitions, we need to examine the first derivative of the entanglement,
\begin{eqnarray}
\frac{\partial M}{\partial T}&=&\frac{\partial M}{\partial S_0}\frac{\partial S_0}{\partial T}=-\frac{\partial M}{\partial S_0}\frac{\partial^2 F}{\partial T^2},\\
\frac{\partial M}{\partial \lambda_i}&=&\frac{\partial M}{\partial \langle H_i\rangle}\frac{\partial \langle H_i\rangle}{\partial \lambda_i}=\frac{\partial M}{\partial\langle H_i\rangle}\frac{\partial^2 F}{\partial \lambda_i^2},i=1,2,\cdots,K.
\end{eqnarray}
These equations show that first derivative of any entanglement is proportional to the second order derivative of free energy. Using entanglement measures, the second order thermal phase transitions can be identified through nonanalytic or vanishing values of $\partial M/\partial T$ at the thermal critical point.

\begin{figure}
\begin{center}
\includegraphics[scale=0.2]{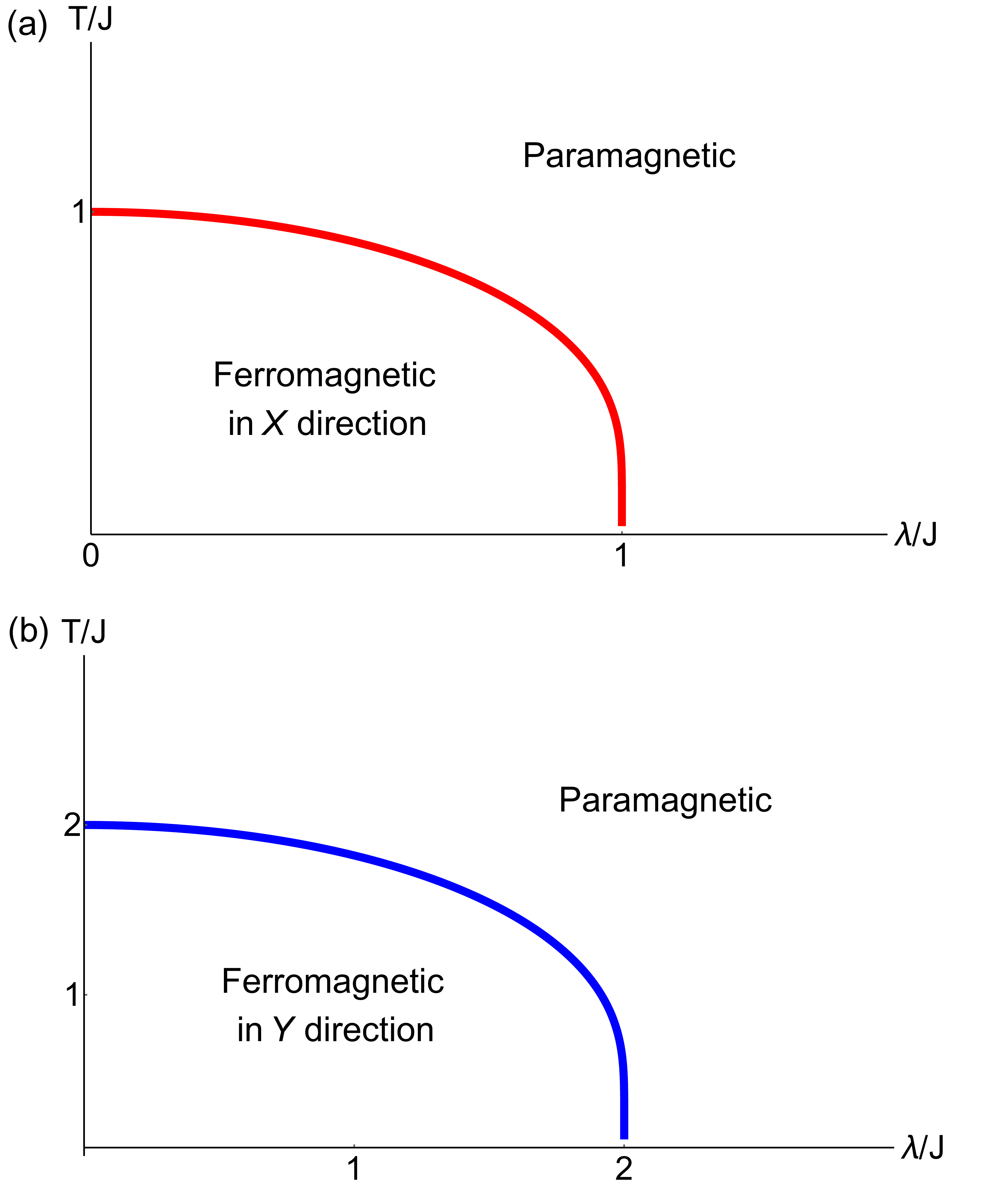}
\end{center}
\caption{(color online). Finite temperature phase diagram of LMG model. (a). Phase diagram at $0<\gamma<1$: At low temperature and weak magnetic field, LMG model is in a ferromagnetic state along $x$ direction. While at high temperature and strong field region, the system is in a paramagnetic phase. The red-solid line is the phase boundary, $T=\lambda/\tanh^{-1}(\lambda)$. (b). Phase diagram at $\gamma>1$: At low temperature and weak magnetic field, LMG model is in a ferromagnetic state along $y$ direction. While at high temperature and strong field region, the system is in a paramagnetic phase. The blue-solid line is the phase boundary, $T=\lambda/\tanh^{-1}(\lambda/\gamma)$. }
\label{fig:epsart1}
\end{figure}

\begin{figure}
\begin{center}
\includegraphics[scale=0.2]{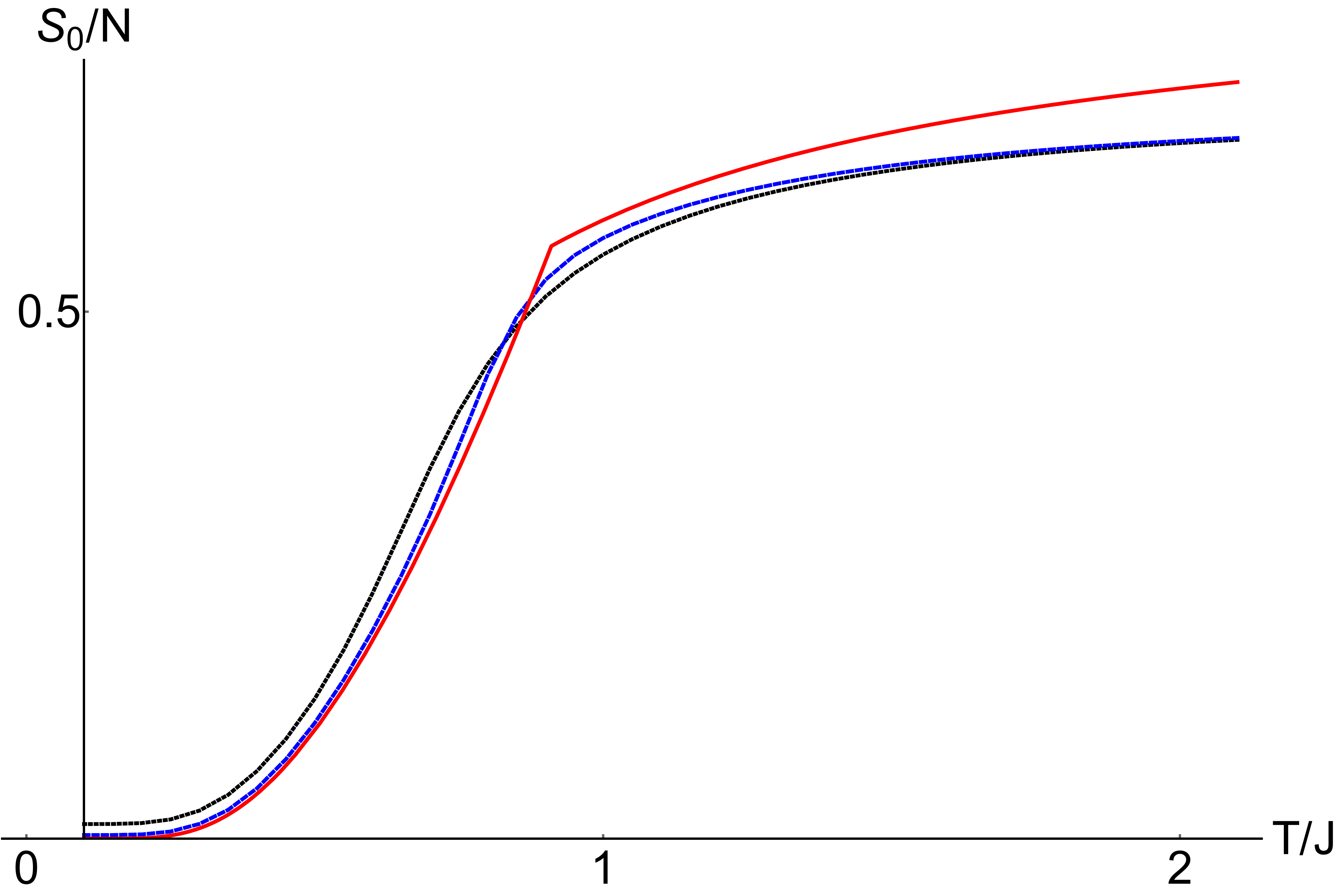}
\end{center}
\caption{Equilibrium entropy per spin $s_0\equiv S_0/N$ as a function of temperature $T/J$ in the LMG model with $\lambda=0.5J$ and $\gamma=0.5$ for $N=50$ (black-dotted line), $N=200$ (blue-dashed line) and $N=\infty$ (red-solid line). }
\label{fig:epsart2}
\end{figure}

\begin{figure}
\begin{center}
\includegraphics[scale=0.25]{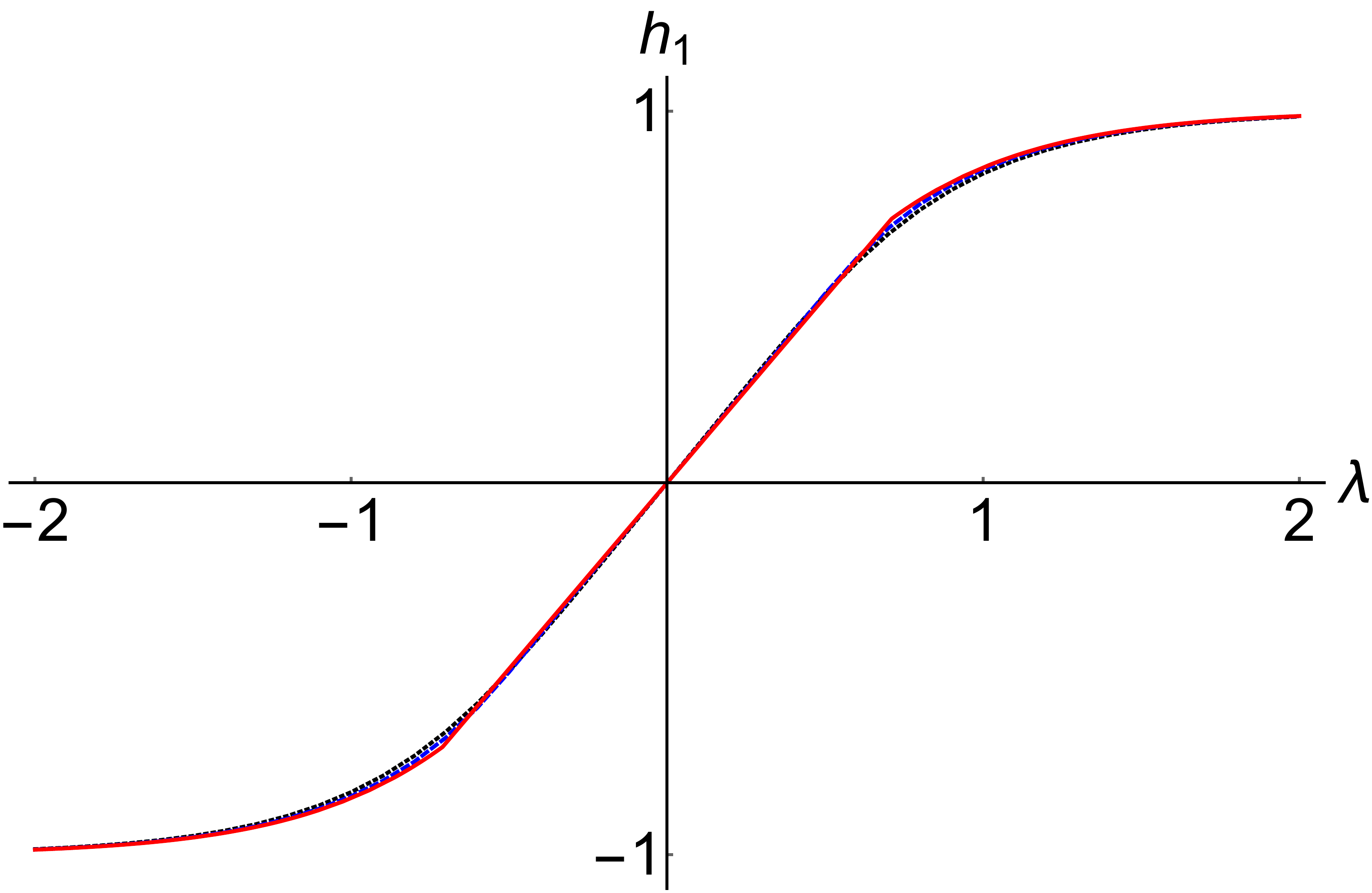}
\end{center}
\caption{(color online). Equilibrium density per spin $h_1\equiv\langle H_1\rangle/N$ as a function of control parameter $\lambda$ in the LMG model with temperature $T=0.8J$ and $\gamma=0.5$ for $N=50$ (black-dotted line), $N=200$ (blue-dashed line) and $N=\infty$ (red-solid line).}
\label{fig:epsart3}
\end{figure}

\section{Model Study}
To demonstrate the above ideas, we study an experimental relevant spin models, namely the  Lipkin-Meshkov-Glick (LMG) model \cite{LMG1965a,LMG1965b,LMG1965c} with Hamiltonian
\begin{eqnarray}
H(\lambda)&=&-\frac{J}{N}\sum_{1\leq i<j\leq N}\Big(\sigma_i^x\sigma_j^x+\gamma\sigma_i^y\sigma_j^y\Big)-\lambda\sum_{i=1}^N\sigma_j^z,\\
&\equiv&H_0+\lambda H_1,
\end{eqnarray}
where $J$ is the ferromagnetic coupling strength between arbitrary two pauli spins $\vec{\sigma}_i$ and $\vec{\sigma}_j$ while $\gamma$ describes the anisotropy of the coupling in the $y$ direction. The LMG model have been experimentally realized in trapped ion systems \cite{Cirac2004,Fried2008,LMGExp2011}.

The LMG model can be simplified by defining a collective spin operator
$S^{\alpha}=\frac{1}{2}\sum_{j=1}^N\sigma_j^{\alpha}$ where
$\alpha=x,y,z$ and it is simple to show that the collective spin operators satisfy the ordinary spin angular
momentum commutation relations. In terms of collective spin operators, the Hamiltonian $H(\lambda)$ can be rewritten as,
\begin{eqnarray}
H_S&=&-\frac{2J}{N}(S_x^2+\gamma S_y^2)-2\lambda S_z+\frac{J(1+\gamma)}{2}.
\end{eqnarray}
It is obvious that the collective spin angular momentum squared is a conserved quantity, $[S^2,H_S]=0$. Thus we can make use the quantum number of the collective spin operator squared $s$ to classify the total Hilbert into different blocks, each of the block take a fixed quantum number of the collective spin angular momentum squared, which can take values $s=\frac{N}{2},\frac{N}{2}-1,\cdots, 1, 0$. Moreover, each collective spin block is highly degenerate and the degeneracy of collective spin-$S$ block is $D(S)=C_N^{N/2-S}-C_N^{N/2-S-1}=C_N^{N/2-S}(2S+1)/(N/2+S+1)$.
By such a mapping, the many-body problem of $H$ in LMG model is simplified into diagonalising a series of small large spins with Hamiltonian $H_S$. Then the partition function of the spin system with Hamiltonian $H$ can be calculated by
\begin{eqnarray}
Z(T,\lambda)&=&\sum_{S=0}^{N/2}D(S)\text{Tr}_S[e^{-\beta H_S(\lambda)}].
\end{eqnarray}
From the partition function we then can get the free energy
\begin{eqnarray}
F(T,\lambda)=-T\ln Z(T,\lambda).
\end{eqnarray}
Thus all the thermodynamic quantities can be evaluated from free energy.

For infinite system, $N\rightarrow\infty$, mean field theory becomes exact \cite{Quan2009} and we show mean field phase diagram of LMG model in Figure 1.  At thermodynamic limit, $N\rightarrow\infty$, when $0<\gamma<1$, we obtain (See Appendix B for derivations) the temperature is a universal function of thermodynamic entropy,
\begin{eqnarray}
T=\left\{
  \begin{array}{ll}
    \frac{\varepsilon^2}{\ln\frac{2}{\sqrt{1-\varepsilon^2}}-\frac{S_0}{N}}, & \hbox{$T<T_c$}; \\
   \frac{\lambda^2}{\ln\frac{2}{\sqrt{1-\lambda^2}}-\frac{S_0}{N}}, & \hbox{$T>T_c$.}
  \end{array}
\right.
\end{eqnarray}
where $\varepsilon=\sqrt{M_x^2+\lambda^2}$. Also the control parameter $\lambda$ is a universal function of the density $\langle H_1\rangle$,
\begin{eqnarray}\lambda=
\left\{
  \begin{array}{ll}
    \frac{\langle H_1\rangle}{N}, & \hbox{$T<T_c$;} \\
    \frac{1}{\beta}\tanh^{-1}\frac{\langle H_1\rangle}{N}, & \hbox{$T>T_c$.}
  \end{array}
\right.
\end{eqnarray}
When $\gamma>1$, the temperature is given in terms of the entropy by
\begin{eqnarray}
T=\left\{
  \begin{array}{ll}
    \frac{\varepsilon^2/\gamma^2}{\ln\frac{2}{\sqrt{1-\varepsilon^2/\gamma^2}}-\frac{S_0}{N}}, & \hbox{$T<T_c$}; \\
   \frac{\lambda^2/\gamma^2}{\ln\frac{2}{\sqrt{1-\lambda^2/\gamma^2}}-\frac{S_0}{N}}, & \hbox{$T>T_c$.}
  \end{array}
\right.
\end{eqnarray}
The control parameter $\lambda$ is given in terms of its density $\langle H_1\rangle$ by
\begin{eqnarray}\lambda=
\left\{
  \begin{array}{ll}
    \frac{\gamma\langle H_1\rangle}{N}, & \hbox{$T<T_c$;} \\
    \frac{1}{\beta}\tanh^{-1}\frac{\langle H_1\rangle}{N}, & \hbox{$T>T_c$.}
  \end{array}
\right.
\end{eqnarray}
These results support our Theorem 2.

In Figure 2, we show the equilibrium entropy per spin $s_0\equiv S_0/N$ in the LMG model at $\lambda=0.5J$ and $\gamma=0.5$ as a function of the temperature for different system sizes, $N=50,200,\infty$. We can see that the entropy is monotonic function of the temperature, which thus supports Theorem 2.

Figure 3 presents the equilibrium density per spin $h_1\equiv\langle H_1\rangle/N$ in the LMG model at temperature $T/J=0.8$ as a function of the control parameter $\lambda$ for different system sizes, $N=50,200,\infty$. We can see that the density $h_1$ is a monotonic function of the control parameter $\lambda$ and it thus supports the Theorem 2.

In Figure 4, we show the behavior of order parameter as a function of entropy, where we use entropy as a fundamental variable. In Figure 4(a), we present $\langle\sigma_j^x\rangle$ as a function of entropy per spin $s_0\equiv S_0/N$ in the LMG model at $N\rightarrow\infty$ with $\lambda=0.5J$ and $\gamma=0.5$. One can see that the order parameter vanishes when entropy approaches a critical value $s_0=0.562335$ which corresponds to the critical temperature $T_c=\lambda/\tanh^{-1}\lambda$ with $\lambda=0.5$. Since the phase transitions in LMG model are of second order, according to Theorem 3, one need to evaluate the first derivative of physical observable. In Figure 4(b), we show $\partial\langle \sigma_j^x\rangle/\partial s_0$ versus $s_0$ in the LMG model at $N\rightarrow\infty$ with $\lambda=0.5J$ and $\gamma=0.5$. One can see that $\partial\langle \sigma_j^x\rangle/\partial s_0$ diverges when entropy approaches a critical value $s_0=0.562335$.

In Figure 5, we use density $\langle H_1\rangle$ as a fundamental variable. In Figure 5 (a), we show the order parameter $\langle \sigma_j^x\rangle$ as a function of density per spin $h_1\equiv\langle H_1\rangle/N$ in the LMG model at $N\rightarrow\infty$ with $T=0.8J$ and $\gamma=0.5$. One can see that the order parameter vanishes as density approaches a critical value $h_1=0.710412$ which corresponds to the critical fields at temperature $T=0.8J$. In Figure 5(b), we present $\partial\langle \sigma_j^x\rangle\partial h_1$ as a function of density per spin $h_1\equiv\langle H_1\rangle/N$ in the LMG model at $N\rightarrow\infty$ with $T=0.8J$ and $\gamma=0.5$. One can see that $\partial\langle \sigma_j^x\rangle/\partial h_1$ diverges when $h_1$ approaches a critical value $h_1=0.710412$.

\begin{figure}
\begin{center}
\includegraphics[scale=0.20]{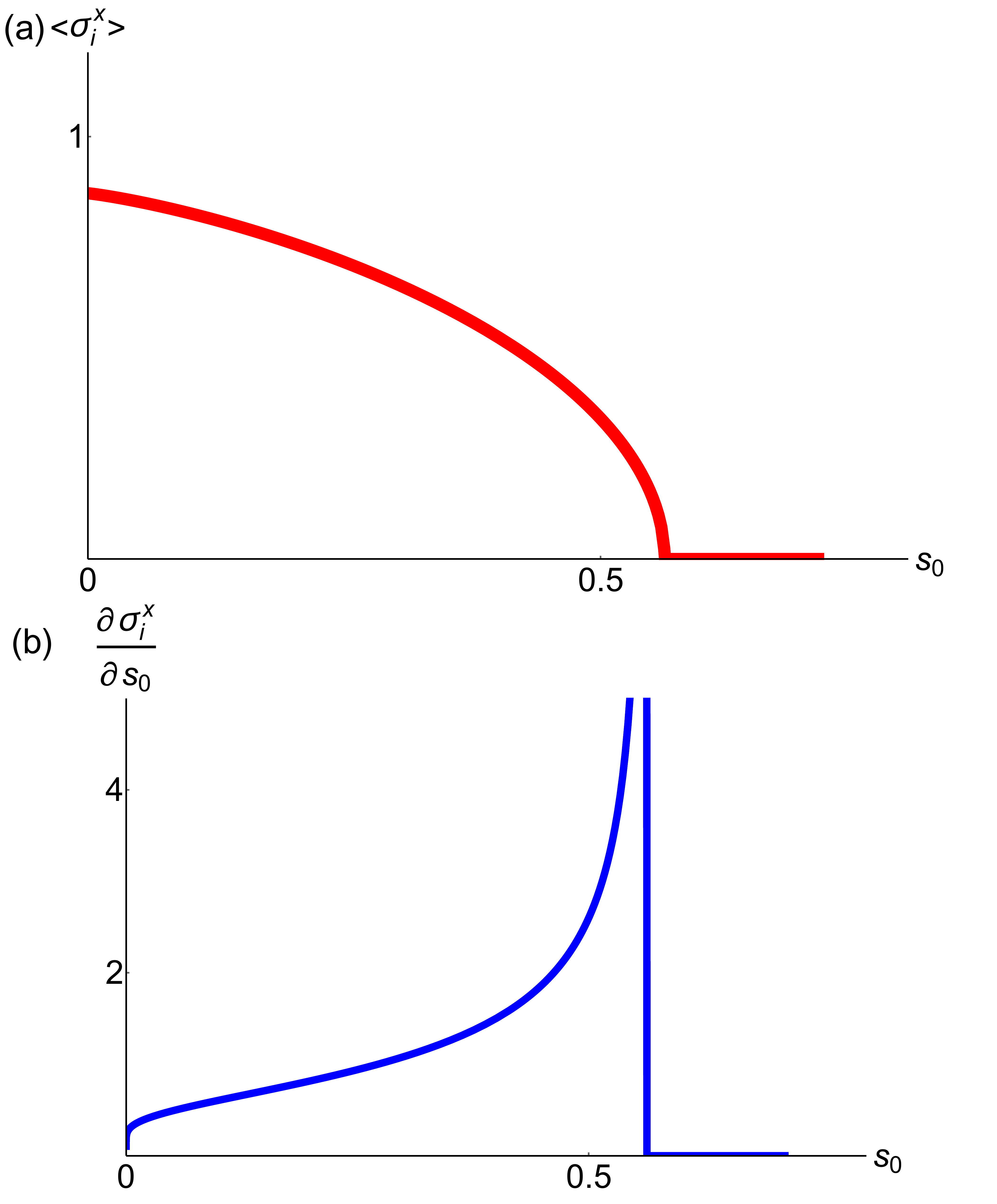}
\end{center}
\caption{(color online). Entropy as a fundamental variable: (a). Order parameter $\langle \sigma_j^x\rangle$ as a function of entropy per spin $s_0\equiv\frac{S_0}{N}$ in the LMG model at $N\rightarrow\infty$ with $\lambda=0.5J$ and $\gamma=0.5$. (b). $\partial\langle \sigma_j^x\rangle/\partial s$ versus $s_0\equiv\frac{S_0}{N}$ in the LMG model at $N\rightarrow\infty$ with $\lambda=0.5J$ and $\gamma=0.5$. The critical value of the entropy per spin is $s=0.562335$ where corresponds to the critical temperature $T_c=\lambda/\tanh^{-1}\lambda$ with $\lambda=0.5$.}
\label{fig:epsart4}
\end{figure}

\begin{figure}
\begin{center}
\includegraphics[scale=0.2]{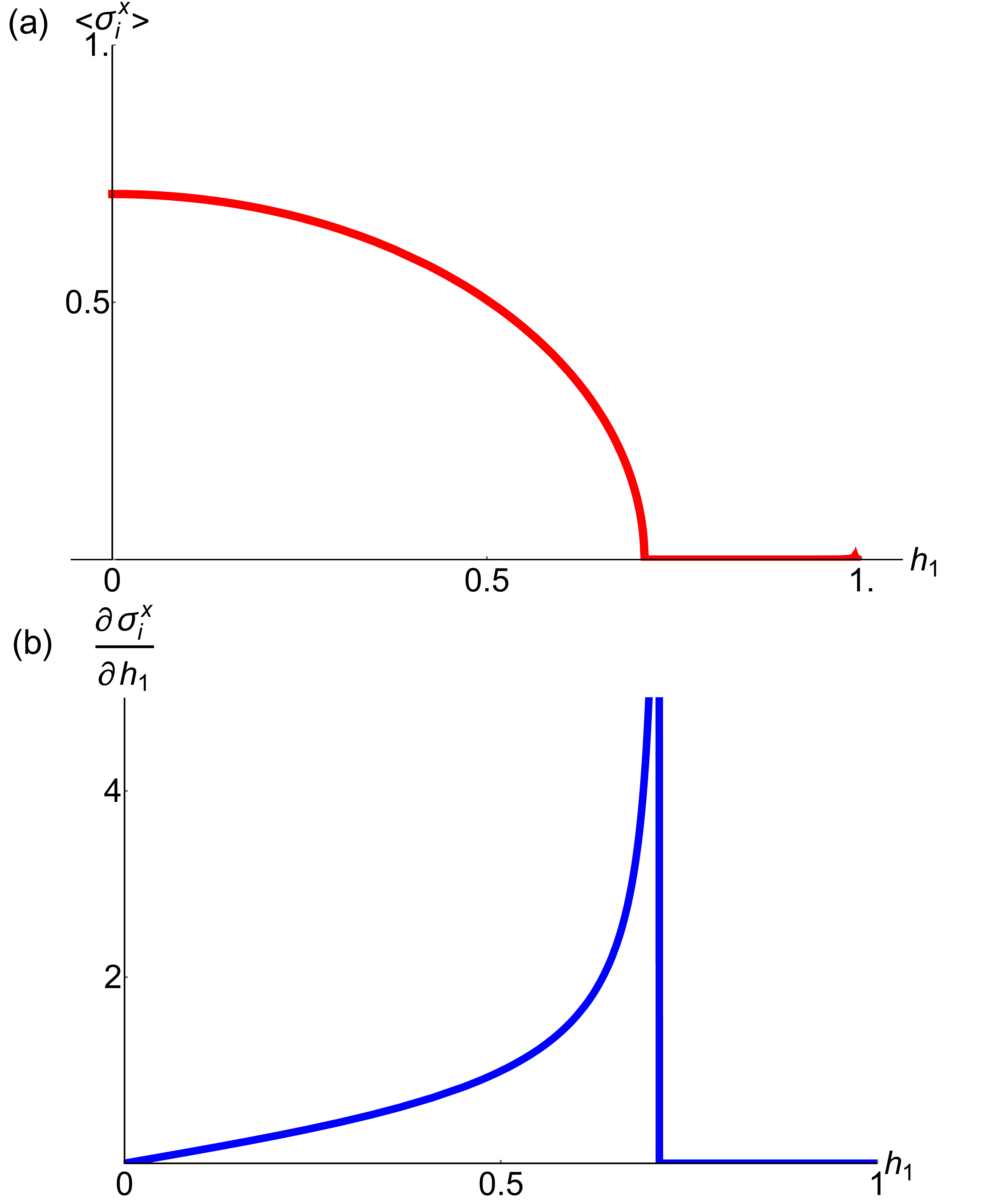}
\end{center}
\caption{(color online). Density as a fundamental variable: (a). Order parameter $\langle \sigma_j^x\rangle$ as a function of density per spin $h_1\equiv\langle H_1\rangle/N$ in the LMG model at $N\rightarrow\infty$ with $T=0.8J$ and $\gamma=0.5$. (b). $\partial\langle \sigma_j^x\rangle\partial h_1$  as a function of density per spin $h_1\equiv\langle H_1\rangle/N$ in the LMG model at $N\rightarrow\infty$ with $T=0.8J$ and $\gamma=0.5$. The critical value of the density is $h_1=0.710412$ which corresponds to the critical fields at temperature $T=0.8J$.}
\label{fig:epsart5}
\end{figure}

Now we evaluate an entanglement measure, namely the concurrence of two spins in the LMG model at zero magnetic field, which is given by (See Appendix C for derivations)
\begin{eqnarray}
M=\max[0,C_1,C_2],
\end{eqnarray}
where
\begin{eqnarray}
C_1&=&|C|-|A|=\langle\sigma_i^x\sigma_j^x\rangle+\langle\sigma_i^y\sigma_j^y\rangle-\langle\sigma_i^z\sigma_j^z\rangle-\frac{1}{4},\nonumber\\
&=&-\frac{1}{4}-\frac{4\langle S^2\rangle-3N}{N(N-1)}-\frac{4}{(N-1)J}\Bigg[F+T\frac{\partial F}{\partial T}+(1-\gamma)\frac{\partial F}{\partial\gamma}\Bigg],\nonumber\\
C_2&=&|D|-|B|,\nonumber\\
&=&\left\{
     \begin{array}{ll}
-\frac{1}{4}+\frac{4\langle S^2\rangle-3N}{N(N-1)}-\frac{4\gamma}{(N-1)J}\frac{\partial F}{\partial\gamma}, & \hbox{$\langle\sigma_i^z\sigma_j^z\rangle<\frac{1}{4}$;} \\
\frac{1}{4}-\frac{4\langle S^2\rangle-3N}{N^2-N}-\frac{4}{(N-1)J}\Big[F+T\frac{\partial F}{\partial T}+\frac{\partial F}{\partial\gamma}\Big], & \hbox{$\langle\sigma_i^z\sigma_j^z\rangle>\frac{1}{4}$.}
     \end{array}
   \right.\nonumber
\end{eqnarray}
We can see that the concurrence is a universal functional of the first derivatives of the free energy, $\partial_{\lambda}F$ and $\partial_{T}F$, this supports our Theorem 4.

\section{Summary}

In summary, we show that density functional theory provide insights for phase transitions and entanglement at finite temperatures. We proved that the equilibrium state of an interacting quantum many-body system which is in thermal equilibrium with a heat bath at a fixed temperature is a universal functional of the first derivatives of the free energy with respect to temperature and other control parameters. This insight from density functional theory enables us to express the average value of any observable and any entanglement measures at nonzero temperature in terms of the first derivatives of the free energy with respect to temperature and other control parameters. These results from density functional theory give new insights for phase transitions and provide new profound connections between entanglement and phase transition in interacting quantum many-body physics.

\begin{acknowledgements}
This work was supported by the startup fundation of Shenzhen University.
\end{acknowledgements}

\section*{Appendix A: Proof the free energy functional is lower bounded by the equilibrium free energy}
 \renewcommand{\theequation}{A\arabic{equation}} \setcounter{equation}{0}
In this appendix we show that the free energy functional
\begin{eqnarray}
F[T,\{\lambda_i\},\rho]=\text{Tr}\Big[\rho\Big(H(\{\lambda_i\})+k_BT\ln\rho\Big)\Big],
\end{eqnarray}
satisfies
\begin{eqnarray}
F[T,\{\lambda_i\},\rho]>F[T,\{\lambda_i\},\rho_0], \rho\neq\rho_0
\end{eqnarray}
for all positive definite density matrices $\rho$ with unit trace and $\rho_0$ is the equilibrium state corresponds to the control parameters $(T,\{\lambda_i\})$.
To prove the inequality, we first define an assistant parameter $g$ dependent density matrix,
\begin{eqnarray}
\rho_{g}\equiv \frac{e^{-\beta[H(\{\lambda_i\})+g O]}}{\text{Tr}[e^{-\beta(H(\{\lambda_i\})+g O)}]},
\end{eqnarray}
with
\begin{eqnarray}
O\equiv-\frac{1}{\beta}\ln\rho-H(\{\lambda_i\}).
\end{eqnarray}
Note that $g$ is an assistant parameter and moreover,
\begin{eqnarray}
\rho_{g=0}&=&\rho_0=\frac{e^{-\beta H(\{\lambda_i\})}}{\text{Tr}[e^{-\beta H(\{\lambda_i\})}]},\\
\rho_{g=1}&=&\rho.
\end{eqnarray}
We want to show that if $\rho\neq\rho_0$,
\begin{eqnarray}
F[T,\{\lambda_i\},\rho]>F[T,\{\lambda_i\},\rho_0].
\end{eqnarray}
For simplicity of notation, we shall not write out explicitly the control parameter dependence in the free energy hereafter as they are fixed. Because
\begin{eqnarray}
F[\rho]-F[\rho_0]&=&\int_0^1d g\frac{\partial F[\rho_g]}{\partial g}.
\end{eqnarray}
To evaluate $\partial_gF[\rho_g]$, we write out
\begin{eqnarray}
F[\rho_g]&=&\text{Tr}\Big[\rho_{g}\Big(H(\{\lambda_i\})+k_BT\ln\rho_g\Big)\Big].
\end{eqnarray}
Thus
\begin{eqnarray}
\partial_gF[\rho_g]&=&\text{Tr}\Big[\partial_g\rho_{g}\Big(H\{\lambda_i\}+k_BT\ln\rho_g\Big)\Big],\\
&=&-g\text{Tr}[O\partial_g\rho_g].
\end{eqnarray}
Using the operator identity \cite{Wilcox1967}
\begin{eqnarray}
\frac{\partial}{\partial g}e^{-\beta (H+g O)}&=&-e^{-\beta (H+g O)}\int_0^{\beta}duO(u),
\end{eqnarray}
where $O(u)\equiv e^{u(H+g O)}Oe^{-u(H+g O)}$, we have
\begin{eqnarray}
\partial_g\rho_g&=&-\int_0^{\beta}du\rho_gO(u)+\rho_g\int_0^{\beta}du\text{Tr}[\rho_gO(u)],
\end{eqnarray}
So
\begin{eqnarray}
\partial_gF[\rho_g]&=&-g\text{Tr}[O\partial_g\rho_g],\\
&=&g\int_0^{\beta}du\text{Tr}[\rho_gO(u)O]-g\text{Tr}[\rho_gO]\int_0^{\beta}du\text{Tr}[\rho_gO(u)],\nonumber\\
&=&g\int_0^{\beta}du\Big[\langle O(u)O\rangle_g-\langle O(u)\rangle_g\langle O\rangle_g\Big],\\
&=&g\int_0^{\beta}du\Big[\langle O(u)O\rangle_g-\langle O\rangle_g^2\Big],\\
&=&g\int_0^{\beta}du\Big[\langle O(u/2)O(u/2)^{\dagger}\rangle_g-\langle O\rangle_g^2\Big],\\
&=&g\int_0^{\beta}du\bigg\langle \Big(O(u/2)-\langle O\rangle\Big)\Big(O(u/2)^{\dagger}-\langle O\rangle\Big)\bigg\rangle_g.
\end{eqnarray}
Note that all the expectation values are performed with respect to $\rho_g$. Therefore we finally have
\begin{eqnarray}
F[\rho]-F[\rho_0]&=&g\int_0^{\beta}du\bigg\langle \Big(O(u/2)-\langle O\rangle_g\Big)\Big(O(u/2)^{\dagger}-\langle O\rangle_g\Big)\bigg\rangle_g,\nonumber\\
&\geq&0.
\end{eqnarray}
The equality holds if and only if $O\propto I$, in this case $\rho=\rho_0$.

\section*{Appendix B: Mean field theory of LMG model}
\renewcommand{\theequation}{B\arabic{equation}} \setcounter{equation}{0}
Here we give the mean field theory of LMG model. We consider the LMG model with Hamiltonian
\begin{eqnarray}
H(\lambda)&=&-\frac{J}{N}\sum_{1\leq i<j\leq N}\Big(\sigma_i^x\sigma_j^x+\gamma\sigma_i^y\sigma_j^y\Big)-\lambda\sum_j\sigma_j^z,
\end{eqnarray}
Defining $M_{\alpha}\equiv\langle\sum_j\sigma_j^{\alpha}\rangle/N, \alpha=x,y,z$, the Hamiltonian can be rewritten as
\begin{eqnarray}
H(\lambda)&=&-\frac{J}{2N}\sum_{i,j}\Bigg[\Big(\sigma_i^x-M_x\Big)\Big(\sigma_j^x-M_x\Big)+\gamma\Big(\sigma_i^y-M_y\Big)\Big(\sigma_j^y-M_y\Big)\nonumber\\
&& +M_x(\sigma_i^x+\sigma_j^x)+\gamma M_y(\sigma_i^y+\sigma_j^y)-M_x^2-\gamma M_y^2\Bigg]-\lambda\sum_j\sigma_j^z.\nonumber\\
\end{eqnarray}
Making mean field approximations and setting the quadratic term vanishes, we get the mean field Hamiltonian
\begin{eqnarray}
H_{MF}&=&-\sum_{j}\Big[M_x\sigma_j^x+\gamma M_y\sigma_j^y-\frac{1}{2}(M_x^2+\gamma M_y^2)+\lambda\sigma_j^z\Big],\nonumber\\
\end{eqnarray}
where we set $J=1$. The mean field Hamiltonian is set of decoupled single Pauli spins and can be easily solved with the following eigenvalues
\begin{eqnarray}
E_{\pm}&=&\pm \sqrt{M_x^2+\gamma^2M_y^2+\lambda^2}\equiv\pm\varepsilon.
\end{eqnarray}
The corresponding eigenvectors are
\begin{eqnarray}
|\psi_+\rangle&=&\frac{-(M_x-i\gamma M_y)|\uparrow\rangle+(E_++\lambda)|\downarrow\rangle}{\sqrt{M_x^2+\gamma^2M_y^2+(E_++\lambda)^2}},\\
|\psi_-\rangle&=&\frac{-(M_x-i\gamma M_y)|\uparrow\rangle+(E_-+\lambda)|\downarrow\rangle}{\sqrt{M_x^2+\gamma^2M_y^2+(E_-+\lambda)^2}}.
\end{eqnarray}
Then the self consistent equations for $M_x,M_y,M_z$ are, respectively
\begin{eqnarray}
M_x&=&\frac{M_x}{\varepsilon}\tanh[\beta\varepsilon],\\
M_y&=&\frac{\gamma M_y}{\varepsilon}\tanh[\beta\epsilon],\\
M_z&=&\frac{\lambda}{\varepsilon}\tanh[\beta\varepsilon].
\end{eqnarray}
Now the phase diagram can be extracted:
\subsection{Physical quantities at $0<\gamma<1$}
For $T<T_c$, the system is in a Ferromagnetic state along $x$ directions, $M_y=0$ and $M_x\neq0$, then the self-consistent equations reduce to
\begin{eqnarray}
\frac{\tanh[\beta\sqrt{M_x^2+\lambda^2}]}{\sqrt{M_x^2+\lambda^2}}&=&1,\\
M_z&=&\lambda.
\end{eqnarray}
The critical temperature $T_c$ is obtained by making $M_x\rightarrow0$, which is
\begin{eqnarray}
T_c=\frac{\lambda}{\tanh^{-1}\lambda}.
\end{eqnarray}
For $T>T_c$, $M_x=M_y=0$, and the self consistent equations reduce to
\begin{eqnarray}
M_z=\tanh\beta\lambda.
\end{eqnarray}
Thus the control parameter is given by
\begin{eqnarray}\lambda=
\left\{
  \begin{array}{ll}
    \frac{\langle H_1\rangle}{N}, & \hbox{$T<T_c$;} \\
    \frac{1}{\beta}\tanh^{-1}\frac{\langle H_1\rangle}{N}, & \hbox{$T>T_c$.}
  \end{array}
\right.
\end{eqnarray}
The entropy per spin for $T<T_c$ is
\begin{eqnarray}
s&=&\frac{S}{N}=-\frac{\varepsilon^2}{T}+\ln\Big[\frac{2}{\sqrt{1-\varepsilon^2}}\Big],
\end{eqnarray}
where $\varepsilon=\sqrt{M_x^2+\lambda^2}$. If $T>T_c$, $\varepsilon=\lambda$, the entropy becomes
\begin{eqnarray}
s&=&\frac{S}{N}=-\frac{\lambda^2}{T}+\ln\Big[\frac{2}{\sqrt{1-\lambda^2}}\Big].
\end{eqnarray}
Thus the temperature is given in terms of entropy by
\begin{eqnarray}
T=\left\{
  \begin{array}{ll}
    \frac{\varepsilon^2}{\ln\frac{2}{\sqrt{1-\varepsilon^2}}-\frac{S}{N}}, & \hbox{$T<T_c$}; \\
   \frac{\lambda^2}{\ln\frac{2}{\sqrt{1-\lambda^2}}-\frac{S}{N}}, & \hbox{$T>T_c$.}
  \end{array}
\right.
\end{eqnarray}

\subsection{Physical quantities at $\gamma>1$}
\noindent For $T<T_c$, the system is in a Ferromagnetic state along $y$ direction, $M_y\neq=0$ and $M_x=0$, then the self consistent equations reduce to
\begin{eqnarray}
1&=&\frac{\gamma\tanh[\beta\sqrt{\gamma^2M_y^2+\lambda^2}]}{\sqrt{\gamma^2M_y^2+\lambda^2}},\\
M_z&=&\frac{\lambda}{\gamma}.
\end{eqnarray}
The critical temperature is obtained by making $M_y\rightarrow0$, which is
\begin{eqnarray}
T=\frac{\lambda}{\tanh^{-1}(\lambda/\gamma)}.
\end{eqnarray}
For $T>T_c$, the system is a paramagnetic state, i.e. $M_x=M_y=0$, the self consistent equations become
\begin{eqnarray}
M_z=\tanh[\beta\lambda].
\end{eqnarray}
Thus the control parameter is given by
\begin{eqnarray}\lambda=
\left\{
  \begin{array}{ll}
    \frac{\gamma\langle H_1\rangle}{N}, & \hbox{$T<T_c$;} \\
    \frac{1}{\beta}\tanh^{-1}\frac{\langle H_1\rangle}{N}, & \hbox{$T>T_c$.}
  \end{array}
\right.
\end{eqnarray}
The entropy per spin for $T<T_c$ is
\begin{eqnarray}
s&=&\frac{S}{N}=-\frac{\varepsilon^2}{\gamma^2T}+\ln\Big[\frac{2}{\sqrt{1-\varepsilon^2/\gamma^2}}\Big],
\end{eqnarray}
where $\varepsilon=\sqrt{M_y^2+\lambda^2}$. If $T>T_c$, $\varepsilon=\lambda$, the entropy becomes
\begin{eqnarray}
s&=&\frac{S}{N}=-\frac{\lambda^2}{\gamma^2T}+\ln\Big[\frac{2}{\sqrt{1-\lambda^2/\gamma^2}}\Big].
\end{eqnarray}
Thus the temperature is given in terms of entropy by
\begin{eqnarray}
T=\left\{
  \begin{array}{ll}
    \frac{\varepsilon^2/\gamma^2}{\ln\frac{2}{\sqrt{1-\varepsilon^2/\gamma^2}}-\frac{S}{N}}, & \hbox{$T<T_c$}; \\
   \frac{\lambda^2/\gamma^2}{\ln\frac{2}{\sqrt{1-\lambda^2/\gamma^2}}-\frac{S}{N}}, & \hbox{$T>T_c$.}
  \end{array}
\right.
\end{eqnarray}

\section*{Appendix C: Derivation of thermal concurrence of two spins in LMG model}
 \renewcommand{\theequation}{C\arabic{equation}} \setcounter{equation}{0}
Here we give the derivation of thermal concurrence of two spins in LMG model. We consider the LMG model at zero magnetic field with Hamiltonian
\begin{eqnarray}
H=-\frac{J}{N}\sum_{i<j}\Big[\sigma_i^x\sigma_j^x+\gamma\sigma_i^y\sigma_j^y\Big],
\end{eqnarray}
To calculate the concurrence, we first reconstruct the two-body reduced density matrix. According to the symmetry of the LMG model, the reduced density matrix for two sites can be written as
\begin{eqnarray}\rho_{ij}=
\left(
  \begin{array}{cccc}
    A & 0 & 0 & D \\
    0 & B & C & 0 \\
    0 & C & B & 0 \\
    D & 0 & 0 & A \\
  \end{array}
\right)
\end{eqnarray}
where
\begin{eqnarray}
A&=&\frac{1}{4}+\langle\sigma_i^z\sigma_j^z\rangle,\\
B&=&\frac{1}{4}-\langle\sigma_i^z\sigma_j^z\rangle,\\
C&=&\langle\sigma_i^x\sigma_j^x\rangle+\langle\sigma_i^y\sigma_j^y\rangle,\\
D&=&\langle\sigma_i^x\sigma_j^x\rangle-\langle\sigma_i^y\sigma_j^y\rangle.
\end{eqnarray}
All the correlation functions can be obtained from free energy
\begin{eqnarray}
\langle\sigma_j^x\sigma_{j+1}^x\rangle&=&-\frac{2}{(N-1)J}\Bigg[F+T\frac{\partial F}{\partial T}-\gamma\frac{\partial F}{\partial\gamma}\Bigg].\\
\langle\sigma_j^y\sigma_{j+1}^y\rangle&=&=-\frac{2}{(N-1)J}\frac{\partial F}{\partial\gamma},\\
\langle\sigma_j^z\sigma_{j+1}^z\rangle&=&\frac{4\langle S^2\rangle-3N}{N^2-N}+\frac{2}{(N-1)J}\Bigg[F+T\frac{\partial F}{\partial T}+(1-\gamma)\frac{\partial F}{\partial\gamma}\Bigg].\nonumber\\
\end{eqnarray}
So the matrix elements of the two-body reduced density matrix is
\begin{eqnarray}
A&=&\frac{1}{4}+\frac{4\langle S^2\rangle-3N}{N^2-N}+\frac{2}{(N-1)J}\Bigg[F+T\frac{\partial F}{\partial T}+(1-\gamma)\frac{\partial F}{\partial\gamma}\Bigg].\nonumber\\
B&=&\frac{1}{4}-\frac{4\langle S^2\rangle-3N}{N^2-N}-\frac{2}{(N-1)J}\Bigg[F+T\frac{\partial F}{\partial T}+(1-\gamma)\frac{\partial F}{\partial\gamma}\Bigg].\nonumber\\
C&=&-\frac{2}{(N-1)J}\Bigg[F+T\frac{\partial F}{\partial T}+(1-\gamma)\frac{\partial F}{\partial\gamma}\Bigg],\\
D&=&-\frac{2}{(N-1)J}\Bigg[F+T\frac{\partial F}{\partial T}+(1+\gamma)\frac{\partial F}{\partial\gamma}\Bigg].
\end{eqnarray}
Then the thermal concurrence is given by \cite{LMGthermal}
\begin{eqnarray}
M&\equiv&\max[0,C_1,C_2],
\end{eqnarray}
where
\begin{eqnarray}
C_1&=&|C|-|A|=\langle\sigma_i^x\sigma_j^x\rangle+\langle\sigma_i^y\sigma_j^y\rangle-\langle\sigma_i^z\sigma_j^z\rangle-\frac{1}{4},\nonumber\\
&=&-\frac{1}{4}-\frac{4\langle S^2\rangle-3N}{N(N-1)}-\frac{4}{(N-1)J}\Bigg[F+T\frac{\partial F}{\partial T}+(1-\gamma)\frac{\partial F}{\partial\gamma}\Bigg],\nonumber\\
C_2&=&|D|-|B|,\nonumber\\
&=&\left\{
     \begin{array}{ll}
-\frac{1}{4}+\frac{4\langle S^2\rangle-3N}{N(N-1)}-\frac{4\gamma}{(N-1)J}\frac{\partial F}{\partial\gamma}, & \hbox{$\langle\sigma_i^z\sigma_j^z\rangle<\frac{1}{4}$;} \\
\frac{1}{4}-\frac{4\langle S^2\rangle-3N}{N^2-N}-\frac{4}{(N-1)J}\Big[F+T\frac{\partial F}{\partial T}+\frac{\partial F}{\partial\gamma}\Big], & \hbox{$\langle\sigma_i^z\sigma_j^z\rangle>\frac{1}{4}$.}
     \end{array}
   \right.\nonumber
\end{eqnarray}


\begin{references}

\bibitem{HK1964}
P. Hohenberg and W. Kohn, Inhomogenous electron gas, Phys. Rev. \textbf{136}, B864(1964).

\bibitem{KS1965}
W. Kohn and L. J. Sham, Self-consistent equations including exchange and correlation effects, Phys. Rev. \textbf{140}, A1133(1965).

\bibitem{DFT2015a}
R. O. Jones, Density functional theory: Its origins, rise to prominence, and future, Rev. Mod. Phys. \textbf{87}, 897 (2015).

\bibitem{DFT2015b}
A. Zangwill, A half-century of density functional theory, Phys. Today, \textbf{68}, 34(2015).

\bibitem{Merin1965}
N. D. Mermin, Thermal Properties of Inhomogeneous Electron Gas, Phys. Rev. \textbf{137}, A1441 (1965).

\bibitem{Runge1984}
E. Runge and E. K. U. Gross, Density-Functional Theory for Time-dependent Systems, Phys. Rev. Lett. \textbf{52}, 997 (1984).

\bibitem{Vedral2008}
L. Amico, R. Fazio, A. Osterloh, V. Vedral, Entanglement in many-body systems, Rev. Mod. Phys. \textbf{80}, 517 (2008).


\bibitem{Wu2005}
L. A. Wu, M. S. Sarandy, D. A. Lidar and L. J. Sham, Linking entanglement and quantum phase transitions from density-functional theory, Phys. Rev. A \textbf{74}, 052335 (2005).


\bibitem{Nagy2013a}
\'{A}. Nagy, M. Calixto and E. Romera, A density-functional view of quantum phase transitions, J. Chem. Theory Comput. \textbf{9}, 1068 (2013).

\bibitem{Nagy2013b}
\'{A}. Nagy and E. Romera, Quantum phase transitions via density-functional theory: Extension to the degenerate cases, Phys. Rev. A \textbf{88}, 042515 (2013).

\bibitem{Jaynes1957a}
E. T. Jaynes, Information Theory and Statistical Mechanics, Phys. Rev. \textbf{106}, 620 (1957).

\bibitem{Jaynes1957b}
E. T. Jaynes, Information Theory and Statistical Mechanics. II, Phys. Rev. \textbf{108}, 171 (1957).


\bibitem{Mutual2013}
J. Iaconis, S. Inglis, A. B. Kallin and R. G. Melko, Phys. Rev. B \textbf{87}, 195134 (2013).


\bibitem{LMG1965a}
H. J. Lipkin, N. Meshkov and A.J. Glick, Validity of many-body approximation methods for a solvable model: (I). Exact solutions and perturbation theory, Nucl. Phys. \textbf{62}, 188 (1965).

\bibitem{LMG1965b}
N. Meshkov, A.J. Glick and H. J. Lipkin, Validity of many-body approximation methods for a solvable model: (II). Linearization procedures, Nucl. Phys. \textbf{62}, 199 (1965).

\bibitem{LMG1965c}
A.J. Glick, H. J. Lipkin and N. Meshkov, Validity of many-body approximation methods for a solvable model: (III). Diagram summations, Nucl. Phys. \textbf{62}, 211 (1965).

\bibitem{Cirac2004}
D. Porras and J. I. Cirac, Effective quantum spin systems with trapped ions. Phys. Rev. Lett. \textbf{92}, 207901 (2004).

\bibitem{Fried2008}
A. Friedenauer, H. Schmitz, J. T. Glueckert, D. Porras and T. Schaetz, Simulating a quantum magnet with trapped ions. Nature Phys. \textbf{4}, 757 (2008).

\bibitem{LMGExp2011}
R. Islam,	E. E. Edwards,	K. Kim,	S. Korenblit,	C. Noh,	H. Carmichael,	G. D. Lin,	L. M. Duan,	C. C. Joseph Wang,	J. K. Freericks	and C. Monroe,
Onset of a quantum phase transition with a trapped ion quantum simulator, Nature comm. \textbf{2}, 377 (2011).


\bibitem{Quan2009}
H. T. Quan and F. M. Cucchietti, Quantum fidelity and thermal phase transitions, Phys. Rev. E \textbf{79}, 031101 (2009).

\bibitem{Wilcox1967}
R. M. Wilcox, Exponential Operators and Parameter Differentiation in Quantum Physics, J. Math. Phys. \textbf{8}, 962 (1967).

\bibitem{LMGthermal}
J. M. Matera, R. Rossignoli, N. Canosa, Thermal entanglement in fully connected spin systems and its RPA description,
Phys. Rev. A \textbf{78}, 012316 (2008).


\end{references}
\end{document}